# A physical neural network training approach toward multi-plane light conversion design

Zheyuan Zhu[1], Joe H. Doerr[2], Guifang Li[1], and Shuo Pang[1]

[1] CREOL, The College of Optics and Photonics, University of Central Florida, 4304 Scorpius St., Orlando, FL 32816-2700.
[2] School of Arts and Sciences, Rutgers, The State University of New Jersey, 94 Brett Road, Piscataway, NJ 08854.

## Abstract

Multi-plane light converter (MPLC) designs supporting hundreds of modes are attractive in high-throughput optical communications. These photonic structures typically comprise >10 phase masks in free space, with millions of independent design parameters. Conventional MPLC design using wavefront matching updates one mask at a time while fixing the rest. Here we construct a physical neural network (PNN) to model the light propagation and phase modulation in MPLC, providing access to the entire parameter set for optimization, including not only profiles of the phase masks and the distances between them. PNN training supports flexible optimization sequences and is a superset of existing MPLC design methods. In addition, our method allows tuning of hyperparameters of PNN training such as learning rate and batch size. Because PNN-based MPLC is found to be insensitive to the number of input and target modes in each training step, we have demonstrated a high-order MPLC design (45 modes) using mini batches that fit into the available computing resources.

## 1. Introduction

Multi-plane light conversion (MPLC) devices are attractive as mode (de)multiplexers in beam shaping and optical communication applications [1]–[4]. As the demand for high optical throughput grows, large-scale mode multiplexing and demultiplexing devices with tens to hundreds of spatial models become increasingly desirable, leading to the continuous search for MPLC designs that push the limit of the number of spatial modes supported in a single device [5]–[8]. Although there has been no consensus on the minimal number of phase masks required for a specific MPLC device, the general rule of thumb suggests that it increases slower than $\mathcal{O}(N)$, where $N$ is the number of spatial mode pairs, given the same requirement on coupling efficiency [9]. Yet when designing a large-scale MPLC system supporting the simultaneous conversion between hundreds of modes, challenges can arise from the high-dimensional parameter space. For example, a 10-mode MPLC using 7 phase masks, each with 256×256 pixels, contains 2 million independent phase parameters, complicating the optimization process.

The conventional wavefront matching (WFM) method transforms the high-dimensional MPLC design process into a low-dimensional sequential optimization problem. WFM iteratively adjusts one mask at a time while fixing the rests to match the wavefronts between forward and backward propagating fields through the photonic structure [10]. As a result, WFM can be considered as a specific optimization sequence in the entire parameter space, leaving other potential optimization processes unexplored. In addition, WFM lacks the flexibility to optimize both phase profiles and distances between masks simultaneously. The WFM results cannot be easily modified to accommodate specific input, output, and inter-mask distances.

In recent years, the state-of-the-art artificial neural network (ANN) models have reached comparable scales as the MPLC models. Motivated by the similarity between optical backward propagation and gradient-based ANN training [8], [11], [12], here we have constructed a physical neural network (PNN) based on the optical propagation model in MPLC. The PNN-based MPLC design leverages the hardware and software development in ANN training [13]–[15] to perform global search in the parameter space, with the capability to adjust various hyperparameters to steer among various optimization paths. This paper is organized as follows. We first modeled the free-space propagation in MPLC with an equivalent PNN and formulated the MPLC design problem as PNN training. We demonstrated that PNN-based design can incorporate various optimization sequences, including the WFM method, which can be expressed as a specific optimization sequence. We then performed PNN training on both phase profiles and distances to refocus an existing MPLC design for different input and output distances, a capability that is not available in WFM. Finally, we explored the MPLC designs generated from tuning the hyperparameters in PNN training, primarily the batch size and learning rate, and discovered that the performances were insensitive to the batch size. This allowed us to break down a large-scale MPLC design into mini batches, a technique commonly used in ANN, to fit the PNN training process into available computing resources without compromising the performance.

## 2. Theory

### 2.1. Multi-plane light conversion (MPLC) model

MPLC transforms a set of orthogonal two-dimensional (2D) input fields into another orthogonal set of target fields with cascaded 2D masks along the longitudinal direction. The masks are typically implemented with phase-only spatial light modulators (SLMs) or fabricated as diffractive optical elements. The spacing between the input fields, phase plates, and output fields are generally on the order of mm. Thus, the electric fields in the MPLC system can be calculated using free-space propagation based on wave optics.

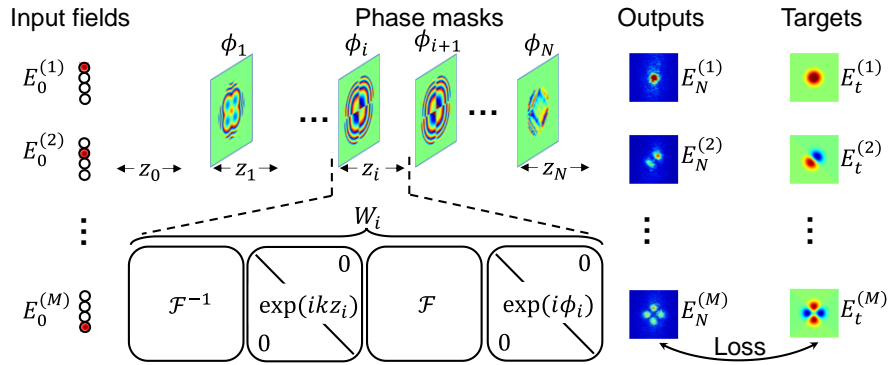

Figure 1. Principle of multi-plane light conversion (MPLC). The system cascades a set of phase masks in free space to successively convert the input fields to target output fields. Each phase mask along with the subsequent free-space propagation is equivalent to a neural network layer. Calculating the optimal masks can be treated as a physical neural network training problem.

Figure 1 illustrates the principle of a light conversion model that aims to transform $M$ input fields into $M$ target fields using $N$ phase plates. The distance between adjacent phase plates $\phi_i$ and $\phi_{i+1}$ is $z_i$. Here we use $z_0$ and $z_N$ to denote the distances from the input fields to the first phase mask, and that from the last phase mask to the output fields, respectively. After discretizing and serializing the fields and masks into vectors, the output field, $E_N$, produced from the input field, $E_0$, can be expressed by successive free-space propagations and phase modulations in matrix form as

$$E_N = \mathcal{F}^{-1}\text{diag}(\exp(ik_z z_N))\mathcal{F}\left[\prod_{i=1}^{N}(\text{diag}(\exp(i\phi_i))\mathcal{F}^{-1}\text{diag}(\exp(ik_z z_{i-1}))\mathcal{F})\right]E_0. \quad (1)$$

Here we have adopted the angular-spectrum method for free-space propagation, where $\mathcal{F}$ and $\mathcal{F}^{-1}$ denote the 2D Fourier transform matrix and its inverse, respectively. $\text{diag}(\mathbf{x})$ creates a diagonal matrix from the vector $\mathbf{x}$. The longitudinal wave vector $k_z = \sqrt{k_0^2 - k_x^2 - k_y^2}$, where $k_x$ and $k_y$ are the spatial frequencies along the x and y dimensions, respectively, and $k_0$ is the wave number of the light in vacuum.

Eq. (1) indicates that the input and output fields follow a series of linear, unitary transforms, defined by the parameters $\phi_i$ and $z_i$. To optimize the parameter sets $\Phi = \{\phi_i, i = 1,2,\ldots,N\}$ and $Z = \{z_i, i = 0,1,\ldots,N\}$ for best mode conversion efficiency, we establish a feed-forward PNN to model the MPLC process with $N+1$ layers, in which the layer $W_i$ consists of the free-space propagation for a distance $z_i$, followed by phase modulation $\phi_i$,

$$W_i = \text{diag}(\exp(i\phi_i))\mathcal{F}^{-1}\text{diag}(\exp(ik_z z_{i-1}))\mathcal{F}, \quad (2)$$

and the last layer, $W_{N+1}$, consists only of free-space propagation for a distance $z_N$,

$$W_{N+1} = \mathcal{F}^{-1}\text{diag}(\exp(ik_z z_N))\mathcal{F}. \quad (3)$$

All the $\phi_i$ and $z_i$ parameters in the layer $W_i$ can be toggled between trainable and untrainable states. For an MPLC design supporting $M$ modes, a set of $M$ output fields $\{E_N^{(j)}, j = 1,2,\ldots,M\}$, generated from a set of $M$ input fields $\{E_0^{(j)}, j = 1,2,\ldots,M\}$ via Eq. (1), are designed to couple into the respective target fields $\{E_t^{(j)}, j = 1,2,\ldots,M\}$. The coupling efficiency, $\eta^{(j)}$, which quantifies the percentage of optical power coupled from output field $E_N^{(j)}$ into the target field $E_t^{(j)}$, is given by the magnitude square of the overlap integral in Eq. (4),

$$\eta^{(j)} = \left|\left(E_N^{(j)}\right)^*\left(E_t^{(j)}\right)\right|^2. \quad (4)$$

Here $*$ denotes the conjugate transpose of a matrix/vector. All pairs of input and target fields are assumed to be normalized, such that $\left|\left(E_0^{(j)}\right)^*\left(E_0^{(j)}\right)\right| = 1$, and $\left|\left(E_t^{(j)}\right)^*\left(E_t^{(j)}\right)\right| = 1$. Since the free-space propagation and phase modulation are both unitary transformations, the output fields are also normalized $\left|\left(E_N^{(j)}\right)^*\left(E_N^{(j)}\right)\right| = 1$.

### 2.2. Optimization in MPLC model

The PNN model given by Eq. (1) is constructed in TensorFlow 2.4, which provides a gradient-based optimizer, Adaptive Moment Estimation (ADAM), to train the entire or a subset of MPLC parameters, including phase profiles on masks $\phi_i$ and the inter-mask distances, $z_i$. The loss function to minimize is the average percentage of power loss, $L$, among all $M$ pairs of output fields $E_N$ and target fields $E_t$,

$$L := \frac{1}{M}\sum_{j=1}^{M}\left(1 - \eta^{(j)}\right) = -\frac{1}{M}\sum_{j=1}^{M}\eta^{(j)} + C. \quad (5)$$

Here we can omit the constant $C$ in the loss function since it is independent on all the MPLC parameters. In addition, we can also designate a subset, $\{j_1, j_2, \ldots, j_B\} \subseteq \{1,2,\ldots,M\}$, of all the input-target field pairs in the loss function as

$$L = -\frac{1}{B}\sum_{b=1}^{B}\eta^{(j_b)}, \quad (6)$$

which is equivalent to using a mini batch of the dataset in neural network training.

The ADAM optimizer moves the parameters according to Eq. (7) for $i = 1,2,...,N$ with considerations on the running averages of the gradient and momentum [16].

$$\phi_i \leftarrow \phi_i - \gamma f(\nabla_{\phi_i} L) \quad (7)$$
$$z_i \leftarrow z_i - \gamma f(\nabla_{z_i} L).$$

Here $f(\cdot)$ summarizes the gradient and momentum scaling operations in each ADAM iteration. $\gamma$ is the learning rate. The gradients, $\nabla_{\phi_i} L$ and $\nabla_{z_i} L$, are calculated by TensorFlow using automatic differentiation.

We can further customize the optimization sequence using macros that define the parameter(s) to fix or update, as well as which pairs of output and target fields to include in the loss, $L$, in each training iteration. For example, we can choose to update the first mask and the distance between output and the last mask in one training step by executing ADAM training (Eq. (7)) only on the parameters $\phi_1$ and $z_N$, while leaving all other parameters fixed. The default PNN optimization sequence executes Eq. (7) on all the masks $\Phi$, while fixing the distances $Z$ between masks.

The conventional wavefront matching (WFM) method is also a gradient-based optimizer with a specific optimization sequence. Contrary to the full parameter update in PNN, WFM considers only one phase mask, $\phi_i$ is trainable in each iteration. By setting the gradient of $L$ with respect to $\phi_i$ to zero, the critical point, $\nabla_{\phi_i} L = 0$, on the coupling efficiency contour satisfies the analytical expression in Eq. (8),

$$\frac{1}{M}\sum_{j=1}^{M}(\xi_{i+1} \exp(i\phi_i)\, \varepsilon_i - c.c.) = 0. \quad (8)$$

Here $c.c.$ denotes the complex conjugate of the preceding term. $\xi_{i+1} = \left(E_t^{(j)}\right)^* \mathcal{F}^{-1}\text{diag}(\exp(ik_z z_N))\mathcal{F}\left[\prod_{k=i+1}^{N} W_i\right]\mathcal{F}^{-1}\text{diag}(\exp(ik_z z_i))\mathcal{F}$ can be interpreted as the conjugate transpose of the backward-propagating field, and $\varepsilon_i = \left[\prod_{k=1}^{i-1} W_i\right] E_0^{(j)}$ can be interpreted as the forward-propagating field, both at the location of mask $i$. WFM chooses $\phi_i$ to match the wavefronts of the forward- and backward-propagating fields, so that Eq. (8) holds.

Figure 2(a) illustrates the optimization sequences undertaken by PNN (Eq. (7)) and WFM (Eq. (8)) to reach their respective solutions. Because of the phase-shift ambiguity, the coupling efficiency contour contains an infinite number of local maxima. WFM jumps to a series of critical points along one of the mask dimensions with the rest of the masks fixed. To reach other maxima, WFM requires different initial conditions, which can be hard to engineer. In contrast, PNN follows the gradient direction towards a local maximum, which can lead to a different solution. In addition to scan the initial conditions, we can change the learning rate and momentum of the training process to access different maxima nearby.

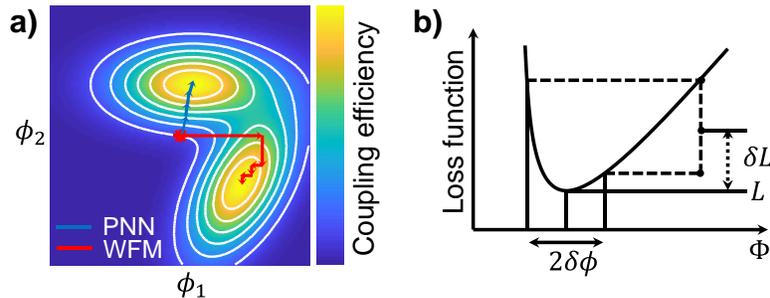

Figure 2. (a) Comparison of the optimization sequences between PNN and WFM on the coupling efficiency contour. (b) Illustration of the evaluation metrics, loss $L$ and sharpness $\delta L$ used in PNN inference.

## 2.3. Evaluation of MPLC model

After optimization, the inference process evaluates both the loss performance $L$, as well as resilience of the model to perturbations using the sharpness metric in ANN [18], defined by the change in the loss function under perturbed phase in Eq. (9),

$$\delta L = \frac{|L_{\Phi+\delta\Phi_k} - L_{\Phi}|}{1 + |L_{\Phi}|}. \tag{9}$$

Here $\delta\Phi$ represents the amount of phase noise added to the MPLC design. $L_{\Phi}$ and $L_{\Phi+\delta\Phi_k}$ denote the average PNN loss given the trained and perturbed phase masks, respectively. $k = 1,2,...,K$ instances of random perturbations are drawn from a uniform distribution $\mathcal{U}_{[-\delta\Phi,\delta\Phi]}$ to calculate the mean and standard deviation of $\delta L$, as illustrated in Figure 2(b). In ANN, a lower sharpness implies the potential of the model to generalize better to the data it has not been trained on before [19]. In PNN-based MPLC model, a smaller sharpness is also preferable, which implies higher optical tolerance against phase errors.

We also evaluate the optical performance using the insertion loss, IL (dB), and optical tolerance $\delta IL$ (dB), which are parallel concepts to PNN loss $L$ and sharpness $\delta L$, respectively. The insertion loss in dB is calculated from the mean eigenvalues of the crosstalk matrix as in Ref. [8]. The optical tolerance is defined as $\delta\text{IL} = |\text{IL}_{\Phi+\delta\Phi_k} - \text{IL}_{\Phi}|$, where $\text{IL}_{\Phi}$ and $\text{IL}_{\Phi+\delta\Phi_k}$ denote the insertion loss (dB) given the trained and perturbed phase masks, respectively. Likewise, $K$ instances of random perturbations are drawn to calculate the mean and standard deviation of $\delta\text{IL}$.

Finally, since masks that differ by a constant phase are identical in terms of functionality, we quantify the similarity, $S$, between two masks, $\phi_1$ and $\phi_2$, using the cross-correlation between complex phasors, $\exp(i\phi_1)$ and $\exp(i\phi_2)$, as defined in Eq. (10),

$$S = \frac{|\int \exp(i\phi_1)\exp(-i\phi_2)\,\mathrm{d}x\mathrm{d}y|}{\sqrt{\int |\exp(i\phi_1)|^2 \mathrm{d}x\mathrm{d}y \int |\exp(i\phi_2)|^2 \mathrm{d}x\mathrm{d}y}}. \tag{10}$$

To evaluate the sharpness and optical tolerance of the MPLC models in the following experiments, we chose $K$=10 random instances with the level of phase perturbation $\delta\Phi$=0.05rad, which is equivalent to 2 quantization levels on a typical 8-bit liquid crystal on silicon (LCoS) SLM.

## 3. Numerical Simulations

### 3.1. Optimization sequences of PNN

We set up a 10-mode MPLC model to compare the phase mask designs from different optimization sequences in PNN. The model was designed to convert a linear array of 10 Gaussian spots into the 10 Hermit-Gaussian modes in the first 4 mode groups using 5 phase plates, shown in Figure 3. The input spots were linearly spaced at 128$\mu$m apart with Gaussian beam waist of 50$\mu$m, matching the output generated from a linear fiber array. The masks contained 512×512 pixels with a pixel size of 3$\mu$m. The distances between the phase masks, as well as the inputs to the first mask, and the output from the last mask, were all 6mm. The output Hermit-Gaussian modes all had beam waists of 200$\mu$m. For comparison with WFM, all the distances were fixed to 6mm as non-trainable parameters. Both WFM and PNN initialized phase masks $\Phi$ with all zeros.

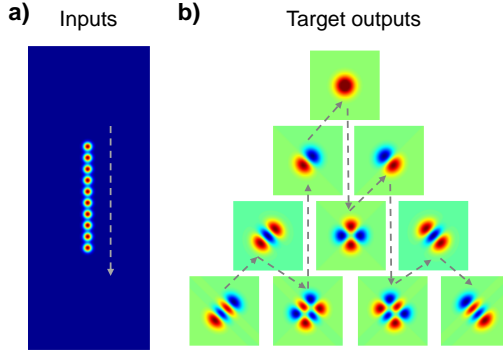

Figure 3. Inputs (a) and target outputs (b) of a 10-mode MPLC model. Arrows indicate the arrangement of the target modes that individual input spots are mapped into.

PNN can incorporate a sequential optimization process using the macro in Table 1, which mimics the behavior of WFM method [17]. The sequential training macro iteratively updates one mask while fixing the rests and the inter-mask distances. Figure 4(a) compares the average coupling efficiency, $\eta$, of all modes as a function of global iteration steps $k$ between WFM and the sequential training macro. The masks after the iterations $k$=1 and 2 are visualized in Figure 4(b). Table 2 lists the similarities, $S$, between the WFM solutions and sequential training macro for each phase mask. The similarities exceed 92% for all phase masks, indicating that WFM and PNN could reach identical solutions.

Table 1: Sequential training macro that matches WFM.

$\varepsilon \leftarrow 10^{-3}, \Phi \leftarrow \mathbf{0}, Z \leftarrow 6\text{mm}$
For $k$ from 1 to $N_{iter}$
    For $i$ from 1 to $N$
        $\delta L \leftarrow 1, L_{prev} \leftarrow 1$
        While $\delta L > \varepsilon$
            Update $\phi_i$ with Eq. (7)
            Calculate loss $L$ with Eq. (5)
            $\delta L \leftarrow |L - L_{prev}|/L_{prev}$
            $L_{prev} \leftarrow L$

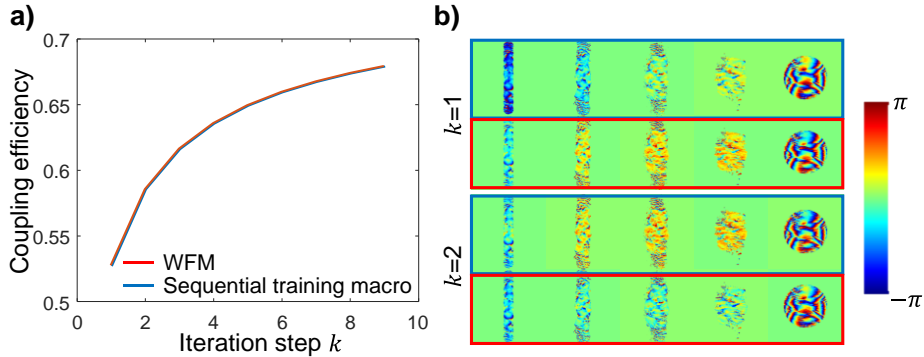

Figure 4. Comparison between the phase mask solutions from wavefront matching and sequential training macro. (a) Average coupling efficiency as a function of iteration steps $k$ for both WFM (red curve) and sequential training macro (blue curve). (b) Phase masks from WFM and sequential macro after the iterations $k$=1 and 2.

Table 2. Similarities between the phase mask solution from WFM and sequential training macro.

| Iteration step $k$ | Mask 1 | Mask 2 | Mask 3 | Mask 4 | Mask 5 |
|---|---|---|---|---|---|
| 1 | 0.9298 | 0.9948 | 0.9921 | 0.9931 | 0.9874 |
| 2 | 0.9701 | 0.9994 | 0.9997 | 0.9998 | 0.9996 |

The default PNN optimization sequence performs the optimizations all the masks. Figure 5(a) compares loss as a function of iteration steps for PNN and WFM. Both algorithms stopped when the relative change in loss function dropped below $10^{-3}$. WFM and PNN converged to average coupling efficiencies of 0.764 and 0.763, respectively. Although PNN took >10 times more steps to reach a similar coupling efficiency as WFM, the overall execution time, 933s, was ~3% shorter than that of WFM, 961s, since TensorFlow can leverage GPUs to parallelize and accelerate the PNN training.

Figure 5(b) compares the five phase masks from WFM and PNN. The similarities between the first to the fifth masks were 0.038, 0.098, 0.077, 0.092, and 0.084, respectively, clearly indicating two different solutions. The sharpness of PNN and WFM solutions are $(7.9 \pm 0.037) \times 10^{-3}$ and $(6.5 \pm 0.014) \times 10^{-3}$, respectively, consistent with the respective optical tolerances, $(1.81 \pm 0.015) \times 10^{-2}$ dB and $(1.81 \pm 0.056) \times 10^{-2}$ dB, of the two designs. These results imply similar local behaviors around the two solutions.

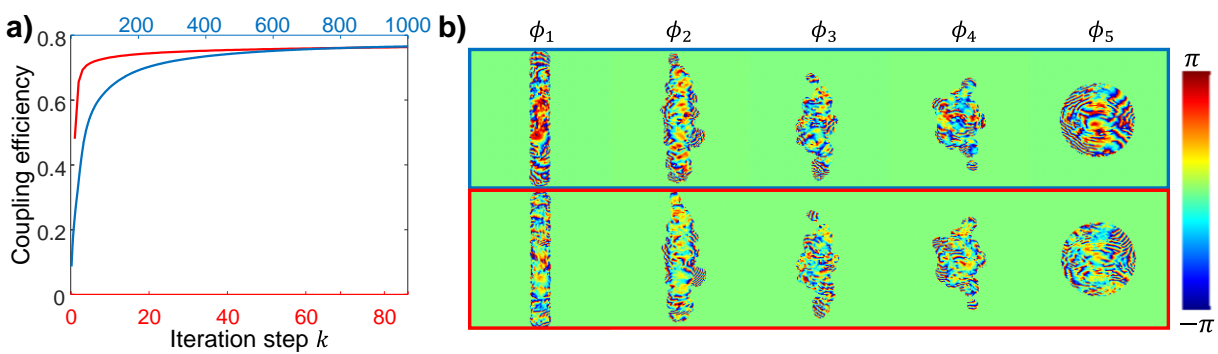

Figure 5. Comparison between the phase mask solutions from WFM and default PNN optimization sequence. (a) Average coupling efficiency as a function of iteration steps $k$ for both WFM (red curve) and PNN (blue curve). (b) Phase masks from WFM and PNN at the end of the iterations.

We have demonstrated that PNN can produce a wider variety of phase mask designs that completely cover the solution space of WFM. Within the PNN framework, WFM can be considered as a special sequential optimization path. It is worth mentioning that PNN is not limited to the sequential (WFM-like) and global training sequences defined in these two examples. Better optimization sequences can be designed to accelerate the convergence of PNN training.

### 3.2. Optimizing inter-mask distances in PNN-based MPLC design

The ability to optimize the input and output distances is an important degree of freedom in MPLC design. Similar to adjusting the object- and image-space distances in lens design [20], tuning the input and output distances alongside the phase masks could potentially improve MPLC performance, or adapt an existing MPLC design to different experimental conditions. However, WFM is not efficient at finding the optimal distances, a task that amounts to enumerating a list of possible MPLC designs with different inter-mask distances, and then executing the WFM iterations from scratch for each configuration. In contrast, the distances and masks are both adjustable in PNN model.

In this example, we optimized the input and output distances $z_0$ and $z_N$ in an existing MPLC design with PNN training. Based on the 5-plate, 10-mode MPLC model, we set the distances $z_0$ and $z_5$, as well as a subset of the phase masks $\Phi$, as trainable parameters. Table 3 summarizes the optimization sequence of the MPLC refocusing macro. In the first round of updates, we trained only the first and last phase masks, $\phi_1$ and $\phi_5$, alongside $z_0$ and $z_5$, which is equivalent to defocusing in conventional lens design. In the second round, we freed all phase masks for training along with $z_0$ and $z_5$. The MPLC model was initialized with the PNN mask design from Figure 5.

Table 3: MPLC refocusing macro that optimizes phase plates alongside distances.

$\varepsilon \leftarrow 10^{-3}, Z \leftarrow 6mm$
Load $\Phi_i$ from Figure 5
// *Round 1: Adjust first and last masks alongside input and output distances.*
$\delta L \leftarrow 1, L_{prev} \leftarrow 1$
While $\delta L > \varepsilon$
    Update $\phi_1$, $\phi_5$, $z_0$, and $z_5$ with Eq. (7)
    Calculate loss $L$ with Eq. (5)
    $\delta L \leftarrow |L - L_{prev}|/L_{prev}$
    $L_{prev} \leftarrow L$
// *Round 2: Adjust all masks alongside input and output distances.*
$\delta L \leftarrow 1, L_{prev} \leftarrow 1$
While $\delta L > \varepsilon$
    Update $\phi_1$, $\phi_2$, $\phi_3$, $\phi_4$, $\phi_5$, $z_0$, and $z_5$ with Eq. (7)
    Calculate loss $L$ with Eq. (5)
    $\delta L \leftarrow |L - L_{prev}|/L_{prev}$
    $L_{prev} \leftarrow L$

Figure 6 plots the phase profiles of the masks after both rounds of updates in the refocusing macro. After round 1 of the updates, the distances $z_0$ and $z_5$ changed from 6mm to 6.34mm and 15.95mm, respectively. Quadratic phase profiles that compensated the curvature of the defocusing wavefront appeared on the last phase mask. After freeing all phase masks alongside $z_0$ and $z_5$, the final design showed a 2.2% improvement in average coupling efficiency, indicating that PNN had performed further optimizations on top of the quadratic phase profile. It is worth noting that the inter-mask distances can also be optimized in a similar way as input and output distances. Constraints can be set to enforce equal distances between adjacent masks to create a feasible reflective MPLC design.

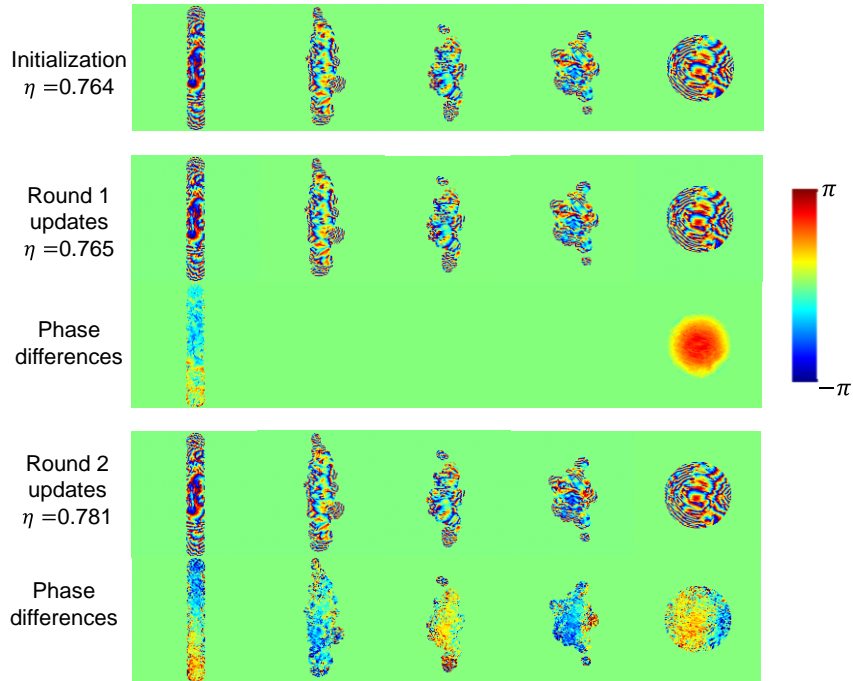

Figure 6. Phase masks after round 1 and round 2 updates with the MPLC refocusing macro.

## 3.3. Effects of batch size in PNN-based MPLC design

Apart from the full MPLC design parameters, PNN also provides access to the hyperparameters for fine-tuning the training process. This example explores the effect of one of the hyperparameters, batch size, on PNN-based MPCL design. Training using mini batches of a full dataset is a common practice in ANN to break down a large dataset into smaller blocks that fit in the available computing resources. Since the pairs of all input and target output electric fields are analogous to the training samples in a full dataset, the potential of using batch training could scale up PNN-based MPCL design to hundreds or thousands of modes.

Table 4: Batch training macro for $M$-mode MPLC with $N$ masks.

$\varepsilon \leftarrow 10^{-3}, \delta L \leftarrow 1, L_{prev} \leftarrow 1$
While $\delta L > \varepsilon$
    Generate random permutations $\{j_1, \ldots, j_M\}$ from $\{1,2,\ldots, M\}$
    $Q \leftarrow \text{ceil}(M/B)$
    For $k=1$ to $Q$
        Calculate loss $L$ with Eq. (6) using $j_{(k-1)B+1}$ to $j_{kB}$
        Update $\phi_1$ to $\phi_N$ with Eq. (7)
    Calculate loss $L$ with Eq. (5)
    $\delta L \leftarrow |L - L_{prev}|/L_{prev}$
    $L_{prev} \leftarrow L$

Table 4 shows the macro to train an $M$-mode MPLC using a batch size of $B$ ($B < M$). The inner loops ensure a full epoch is consumed before updating the loss function of the MPLC model. We applied the batch training macro to the 10-mode example with a batch of 4, 6, 8, and 10 pairs of input and output electric fields in during the parameter update (Eq. (7)). Here a batch size of 10 is equivalent to training using the full dataset. For each batch size, we tuned the learning rate from 0.1 to 0.9 with a step size of 0.1 and selected the resulting MPLC models with the best PNN loss.

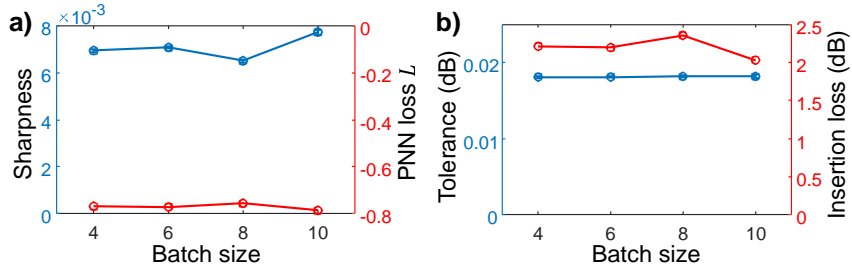

Figure 7. PNN design of the 10-mode MPLC with batch training. (a) Sharpness (blue) and ANN loss $L$ (red) with respect to the training batch size. (b) Insertion loss (red) and tolerance (blue) after phase perturbation as a function of batch size.

Figure 7 plots the loss and sharpness with respect to different batch sizes. The different batch sizes introduced marginal changes in PNN loss, which were comparable in order of magnitude as the changes from phase perturbation. In contrast, there has been considerable evidence indicating that changing the batch size often results in the change in sharpness in conventional ANN [19], [21]. We attribute this behavioral difference to the lack of nonlinearities in MPLC (see Appendix), as the electric field propagation from one phase plate to the next can be completely described by a linear transformation (Eq. (2)). Since batch size does not affect the performance or tolerance of a linear PNN model, we can expand PNN to a high-order MPLC design, using mini batches of the input and target mode pairs.

Figure 8(a) shows a 45-mode MPLC example we have used to test PNN training with mini batches. The model was designed to convert a linear array of 45 Gaussian spots into the 45 Hermit-Gaussian modes in the first 9 mode groups using 8 phase plates. The input spots were linearly spaced at 127$\mu$m apart with Gaussian beam waist of 30$\mu$m. The masks contained 1280×512 pixels with a pixel size of 5μm. The total number of trainable parameters were 5.2 million. The distances between the phase masks, as well as the inputs to the first mask, and the output from the last mask, were fixed at 24mm. The output Hermit-Gaussian modes all had a beam waist of 200$\mu$m. We performed the PNN training using batch sizes of 4, 6, 8, and 45 (full dataset). Within each batch size, we tuned the learning rate from 0.1 to 0.9 with a step size of 0.1 and selected the resulting MPLC models with the best PNN loss.

Table 5: Full-dataset update macro for $M$-mode MPLC with $N$ masks.

$B \leftarrow 8$, $\varepsilon \leftarrow 10^{-3}$, $\delta L \leftarrow 1$, $L_{prev} \leftarrow 1$
While $\delta L > \varepsilon$
    Generate random permutations $\{j_1, \ldots, j_M\}$ from $\{1, 2, \ldots, M\}$
    $Q \leftarrow$ ceil $(M/B)$
    Initialize $\{g_i\}$ as zero vectors
    For $k=1$ to $Q$
        Calculate loss $L$ with Eq. (6) using $j_{(k-1)B+1}$ to $j_{kB}$
        For $i=1$ to $N$
            // Aggregate partial gradients
            $g_i \leftarrow g_i + \nabla_{\phi_i} L$
    Update $\phi_1$ to $\phi_N$ with Eq. (7)
    Calculate loss $L$ with Eq. (5)
    $\delta L \leftarrow |L - L_{prev}|/L_{prev}$
    $L_{prev} \leftarrow L$

Given the size of the model and the number of trainable parameters, the GPUs on our host PC (two NVIDIA RTX2080Ti) supported a maximum batch size of $B=8$. To test the MPLC design trained with full dataset, we wrote the macro in Table 5 to perform an equivalent full-dataset training. Figure 8(b) and (c) compare the loss and sharpness with respect to the batch sizes. All models reached >71% average coupling efficiency and <3dB insertion loss. Considering the variance of the models due to phase perturbation, no noticeable change in the sharpness and optical tolerance can be observed across different batch sizes, indicating successful training of 45-mode MPLC using smaller batch sizes. Hence, we can divide the entire mode-conversion pairs into mini batches that fit in the available memory of individual CPU cores and GPUs, making the PNN training applicable to large-scale MPLC models supporting hundreds to thousands of modes.

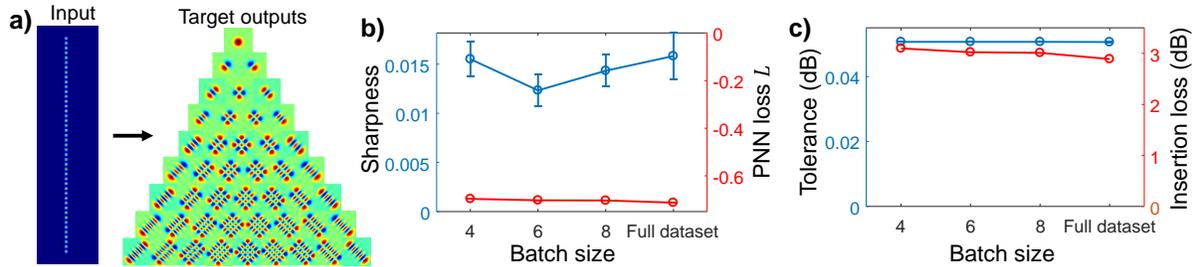

Figure 8. PNN design of a 45-mode MPLC with batch training. (a) Inputs and target outputs of PNN. (b) Sharpness (blue) and ANN loss $L$ (red) with respect to the training batch size. (c) Insertion loss (red) and tolerance (blue) after phase perturbation as a function of batch size.

## 4. Summary

We have demonstrated a new approach towards MPLC design based on a PNN model that incorporates optical propagation and phase modulation. The proposed method can perform simultaneous searches over all the design parameters, including the phase profiles and distances between the masks. The PNN training opens the possibility for engineering the underlying optimization sequences, constituting a superset of the conventional WFM design methods. We have demonstrated the capability to optimize input and output distances alongside the masks, yielding a superior solution that cannot be easily reached by WFM. Combined with the hyperparameter tuning techniques, we have arrived at solutions with similar performances but using smaller batch sizes. The ability to perform batch training could scale up PNN-based MPLC design approach to hundreds or thousands of modes while maintaining compatibility with limited computing resources.

## Acknowledgement

This work is supported in part by the National Science Foundation (ECCS-1932858), and the Office of Naval Research (N00014-20-1-2441). We would like to thank Dr. Stephen Becker from University of Colorado at Boulder for helpful discussions.

## Appendix

Supported by universal approximation theorem [22], ANN models often contain nonlinear activation functions to ensure sufficient expressive power. We speculate that the inclusion of nonlinear activations in conventional ANN contributes to the tradeoff between batch size and sharpness, which we did not observe in linear MPLC model. Here we provide empirical evidence supporting our hypothesis using a fully connected MNIST classification model with two hidden layers. The model has an input of size 784, 2 hidden layers with ReLU of size 100, and an output layer of size 10. We quantify the number of activations as the average number of hidden neurons inhibited by ReLU over all input samples in the dataset. The model was compiled with cross entropy loss and trained with Adam optimizer using batch sizes of 600 and 6000. For the batch size 6000, we aggregated the partial gradient within 10 minibatches of 600 before performing the weight update, as described by the full parameter update macro in Table 5.

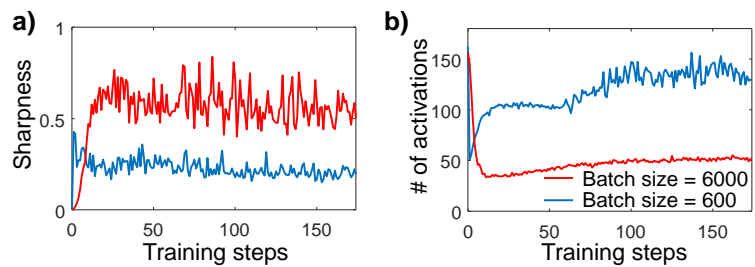

Figure 9. Training with different batch sizes in MNIST model. (a) Sharpness and (b) number of activations as a function of the training steps for batch size of 600 (blue curves) and 6000 (red curves).

We logged the sharpness and the number of activations during the training process. Figure 9 shows that small batch training leads to a lower sharpness than large batch training, consistent with the observations in Ref.[19]. The lower sharpness can be explained by the larger number of activations in small-batch training as when more neurons are inhibited, the outputs are less likely to be impacted by perturbation. The correlation between activations and sharpness has implications towards improvements to ANN generalization algorithms as it provides a novel metric to consider. Further understanding of why batch size influences the number of activations could be valuable in controlling the number of activations in the future as well as using large batch training without the cost of generalization capabilities. In the absence of activation functions in the MPLC model, batch size hardly affects the sharpness and optical tolerance.